\documentclass{iopart}
\usepackage[utf8]{inputenc}
\usepackage{graphicx}
\usepackage{float}
\usepackage{subcaption}
\usepackage{xcolor}
\usepackage{dcolumn}
\usepackage{bm}
\usepackage{tabularx}
\usepackage{booktabs}
\usepackage{amsmath}
\usepackage{amssymb}
\usepackage{amsfonts}
\usepackage[colorlinks=true,linkcolor=blue,urlcolor=blue,citecolor=blue]{hyperref}
\usepackage{multirow}

\newcommand{\bfr}{\bm{r}}

\newcommand{\s}{_\mathrm{{\scriptscriptstyle S}}}
\newcommand{\h}{_\mathrm{{\scriptscriptstyle H}}}
\newcommand{\xc}{_\mathrm{{\scriptscriptstyle XC}}}

\newcommand{\ext}{_\mathrm{{\scriptscriptstyle ext}}}

\newcommand{\Eqref}[1]{Eq.~(\ref{#1})}
\newcommand{\pde}{_\mathrm{{\scriptscriptstyle PDE}}}

\begin{document}

\title{Inverting the Kohn-Sham equations with physics-informed machine learning}

\author{Vincent Martinetto$^{1,*}$, Karan Shah$^{2,3,*}$, Attila Cangi$^{2,3,\dag}$, Aurora Pribram-Jones$^{1,\ddag}$}
\address{$^1$ Department of Chemistry and Biochemistry, University of California Merced, 5200 North Lake Rd., Merced, California 95343, USA}
\address{$^2$ Center for Advanced Systems Understanding, 02826 G\"orlitz, Germany}
\address{$^3$ Helmholtz-Zentrum Dresden-Rossendorf, 01328 Dresden, Germany}
\address{$*$ these authors have contributed equally.}
\address{$\dag$ a.cangi@hzdr.de}
\address{$\ddag$ apj@ucmerced.edu}

\vspace{10pt}
\begin{indented}
\item[]\today
\end{indented}

\maketitle
\ioptwocol

\section{Abstract}
Electronic structure theory calculations offer an understanding of matter at the quantum level, complementing experimental studies in materials science and chemistry. One of the most widely used methods, density functional theory (DFT), maps a set of real interacting electrons to a set of fictitious non-interacting electrons that share the same probability density. Ensuring that the density remains the same depends on the exchange-correlation (XC) energy and, by a derivative, the XC potential. Inversions provide a method to obtain exact XC potentials from target electronic densities, in hopes of gaining insights into accuracy-boosting approximations. Neural networks provide a new avenue to perform inversions by learning the mapping from density to potential. In this work, we learn this mapping using physics-informed machine learning (PIML) methods, namely physics informed neural networks (PINNs) and Fourier neural operators (FNOs). We demonstrate the capabilities of these two methods on a dataset of one-dimensional atomic and molecular models. The capabilities of each approach are discussed in conjunction with this proof-of-concept presentation. The primary finding of our investigation is that the combination of both approaches has the greatest potential for inverting the Kohn-Sham equations at scale.

\section{Introduction}
Modern, high-performance computational resources have enabled large-scale electronic structure simulations of molecules, materials, and other systems of interest across biology, chemistry, physics, and beyond. Kohn-Sham (KS) Density Functional Theory (DFT)~\cite{HK64,KS65} is the most widely used method due to its accuracy and computational efficiency. KS-DFT has helped solve major scientific and technological problems such as simulating chemical reactions, computing material properties, finding new catalysts, discovering drugs, and modeling microscopic environmental processes.  

The electronic structure problem is typically tackled by solving the non-relativistic Schrödinger equation for the molecular Hamiltonian within the Born-Oppenheimer approximation. KS-DFT simplifies this process by transforming the many-body problem into an effective single-particle problem, utilizing the one-to-one correspondence of the external potential and the electronic density~\cite{HK64}. The crux of KS-DFT is producing an electronic density that matches the one obtained from solving the interacting many-body problem~\cite{KS65}. This is accomplished through an effective single-particle potential known as the KS potential, which is comprised of the external potential, the Hartree potential that accounts for electrostatic repulsion, and an exchange-correlation (XC) potential compensating for all interactions beyond the Hartree potential.

While an exact expression for the XC potential ($v\xc=\delta E\xc/\delta n$) is not known, useful approximations exist. These approximations are categorized based on increasing complexity and accuracy~\cite{PS01}. They include, for example, the local density approximation (LDA)\cite{KS65,LYP88,PW92}, which is derived from the interacting uniform electron gas~\cite{CA80}; generalized gradient approximations (GGAs)\cite{PBE96,B88,LYP88,PRCV08}, relying on density gradients; meta-GGAs~\cite{TPSS03,SRP15}, which consider the kinetic energy density; and hybrid functionals, \cite{B93,LYP88,AB99} incorporating a dependence on Hartree-Fock orbitals. More sophisticated approximations beyond hybrid functionals include the random phase approximation\cite{LP77,F01,FG02,HK08,FV05,CVA17} and double hybrid\cite{G06,TKS08,KTL08,MS20} methods.

Despite the usefulness of existing approximations for various applications in chemistry and materials science, developing more accurate XC approximations is an active area of research.~\cite{PGB14} New paths for finding and optimizing such approximations are required for describing complex molecular systems of upcoming interest~\cite{MBSP2017}. Some advances in functional construction leverage machine learning (ML) techniques~\cite{FSBC22,PKK22}. These methods share the common goal of learning the XC energy functional from accurate data on molecular systems~\cite{LWD17,LM19,CZW21,MR21,NTS19,SBCM19,SNH20,MSCN20,BVL20,BZSK20,DF20,FKS20,NAS20,MV20,LK20,SS19,ZVSI19,PDC19,HLBB18,VCDH18,MQG20,NAS22,BK22,KPC22}.

Another promising approach to constructing approximations to the XC functional is by reverse-engineering the XC potential from systems that can be solved exactly. This procedure takes the electronic density as input and finds the corresponding XC potential by inverting the KS equations. Though straightfoward to describe, this process is an example of an inverse problem. Inverse problems are encountered in various scientific fields, such as image reconstruction, a fundamental method in MRI and CT scans; inverse scattering, commonly used in seismology and radar imaging; and tomography, frequently employed in CT scans. In each case, as in our case, observed data is used to reconstruct the structure of an object that gives rise to the initial data~\cite{S00}.
Inverse problems are often ill-posed, meaning they are highly sensitive to small changes in the observed data and noise. This sensitivity makes them challenging to solve~\cite{K22}.

Various algorithmic and numerical schemes for inverting the KS equations exist in the literature~\cite{WP93,JW18,CLG20,ZMP94,WY03,WSBW12,LB94}. These have provided valuable insights\cite{YTPB14,EH14,LEM13,EFRM12,DLM22,GSH21} into the properties of the XC potential, such as the appearance of the derivative discontinuity or the dynamical step structures in time-dependent DFT, and have guided the construction of XC approximations~\cite{CMW08}. However, traditional inversion methods are often plagued by the numerical errors and instabilities~\cite{JW18} common in the solution of inverse problems, as well as from those specific to the density-to-potential problem.

Recent research has demonstrated numerous examples of utilizing ML, particularly neural networks, to address inverse problems, with a focus on medical image processing applications. The primary objective is to learn stable mappings that can be effectively applied to noisy data in a general context~\cite{WEM20,GMM23}. The fusion of physics with ML offers innovative solutions to complex problems, leading to the evolution of physics-informed machine learning (PIML) techniques \cite{karniadakisPhysicsinformedMachineLearning2021a}. Instead of solely relying on data, PIML incorporates governing physical laws, typically differential equations, into the learning process. This results in data-efficient learning and equips the model with the capability to generalize within physically consistent bounds. This is especially relevant in cases with sparse or noisy data where conventional ML methods might not perform well due to overfitting or by producing non-physical outcomes.

In this work, we employ two recent advances in PIML, namely physics-informed neural networks (PINNs)\cite{RPK19} and Fourier neural operators (FNOs)\cite{liFourierNeuralOperator2020}, to tackle the inverse problem in KS-DFT. 
PINNs have the unique ability to learn from the underlying physics described by partial differential equations, leading to improved generalizability with less data and training time compared to conventional methods. This characteristic proves highly advantageous in computational sciences where data availability is limited and generating extensive training sets is costly. PINNs have demonstrated their effectiveness and versatility in various scientific domains for inversions, including materials and fluid dynamics models~\cite{YMK21,CLK20,LMK21}.

Neural operators (NOs) are a class of models that map function-to-function spaces, as opposed to the finite-dimensional vector space mappings of neural networks. This is especially useful for inverse PDE problems where experimental or simulation data is available. FNOs are a type of NO represent operator weights in Fourier space. The main advantages of FNOs are that they generalize well across function spaces, and being resolution-invariant, they can be used for inference on higher resolution grids than the training set grids. FNOs have also been used for forward and inverse PDE problems in various domains \cite{liFourierNeuralOperator2020,kovachkiNeuralOperatorLearning2023a}.

In the following, we construct both a PINN and an FNO to predict the XC potential based on input electronic densities. We demonstrate the implementation of these computational workflows and compare both approaches in terms of efficiency and accuracy for their applicability as KS inverters. We observe higher levels of accuracy and ease of use for FNOs and more predictable errors and greater extrapolation power for PINNs.

\section{Methods}

\subsection{Kohn-Sham density functional theory}
In the framework of KS-DFT, we solve a set of effective single-particle equations known as the KS equations (which we present here using atomic units\cite{Englert1988} for clarity and simplicity of notation), 
\begin{align} \label{KSeqn}
\left[ -\frac{\nabla^2}{2} + v_{s}(\bfr) \right] \phi_i(\bfr) = \epsilon_i ~ \phi_i(\bfr)\,,
\end{align}
where $\phi_i(\bfr)$ represents the KS eigenfunctions and $\epsilon_i$ denotes the KS eigenvalues.
These KS equations constitute an auxiliary system designed to replicate the electron density of an interacting many-body system, expressed as
\begin{align} \label{denseqn}
n(\bfr) = \sum_{i=1}^{N} \vert \phi_i(\bfr) \vert^2\,,
\end{align}
where the summation index corresponds to the total number of electrons, denoted as $N$.

The correspondence of the density defined in Eq.~(\ref{denseqn}) with the density of an interacting many-body system is established through the KS potential which mimics the inter-electronic interaction and is defined as 
\begin{align}
v\s(\bfr) = v\ext(\bfr) + v\h(\bfr) + v\xc(\bfr)\ .
\end{align}
Here, $v\ext(\bfr)$ is the external potential, $v\h(\bfr)$ is the Hartree potential (i.e., the electrostatic potential due to an electronic density interacting with itself) and $v\xc(\bfr)$ is the XC potential. Both the external and Hartree potentials are known exactly in terms of the density. The only unknown is the XC potential. While solving the set of KS equations with the exact XC potential yields the exact density of the many-body system, in practice, approximations to the XC energy need to be used. A plethora of approximations is available in the literature~\cite{KS65,LYP88,PW92,CA80,PBE96,B88,PRCV08,TPSS03,SRP15,B93,AB99} and more accurate approximations are continuously being developed. 

Usually, the KS equations are solved iteratively. This means that a system is defined and an XC approximation is selected. An initial guess of the ground-state density is made, which in turn provides guesses for the Hartree and XC potentials. \Eqref{KSeqn} and \Eqref{denseqn} are solved, generating a new density, which can be used to calculate new Hartree and XC potentials. This cycle will continue until the density converges to within a chosen criterion. In the end, the resulting density is an accurate approximation of the interacting many-body system, which can be used to calculate electronic properties.

Conversely, \Eqref{KSeqn} and \Eqref{denseqn} can be solved to find the exact KS potential that results in a given target density. In this case, no approximation is made to the XC potential, only an initial guess. Then, by solving \Eqref{KSeqn} and \Eqref{denseqn}, the guess for the XC potential can be updated. This process is much less stable than the ``direct" one described above because the potential is sensitive to small changes in the density. 

We use the iDEA code \cite{hodgsonExactTimedependentDensityfunctional2013a, IDEAInteractingDynamic2023} for generating data and as benchmark for inversion. In iDEA, the density-to-potential inversion is done iteratively:
\begin{align}
v_{s} \rightarrow v_{s} + \mu [n(\bfr)^p - n_0(\bfr)^p]\,,
\end{align}
where $n_0(r)$ represents the target density, while $\mu$ and $p$ serve as convergence parameters.

\subsection{Model system}
\begin{figure}
    \centering
    \includegraphics[scale=.5]{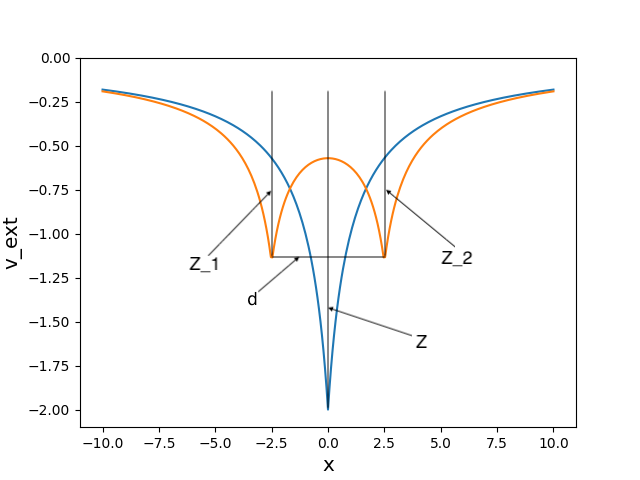}
    \caption{Illustration of the model systems: one-dimensional atom (blue) and a one-dimensional diatomic molecule (orange) with variable parameters charge $Z$, $Z_1$, and $Z_2$ and bond length $d$.}
    \label{fig:ext_pot}
\end{figure}

Model systems in one dimension are a useful tool for the development of DFT approximations~\cite{WSBW12} for two reasons. First, it is simpler to obtain highly accurate or exact data from a one-dimensional system. Second, these model systems can be finely tuned to mimic a wide range of well-understood physical phenomena. Therefore, we leverage one-dimensional models to demonstrate the utility of our inversion method.
We define two model systems: a one-dimensional atom and a one-dimensional diatomic molecule, as illustrated in Figure~\ref{fig:ext_pot}. 
A one-dimensional atom is defined by the external potential
\begin{align}
    v\ext(\bfr) = -\frac{Z}{|\bfr|+a}\,,
\end{align}
where $Z$ is the charge of the atomic well and $a$ is softening parameter to ensure that the potential is defined at all points in space.
This simple model can readily be extended to a one-dimensional model of a diatomic molecule by incorporating two potential wells through an additional parameter, $d$, which represents the bond length in terms of the separation distance between the two wells:
\begin{align}
    v\ext(\bfr) = -\frac{Z_1}{|\bfr-\frac{d}{2}|+a} - \frac{Z_2}{|\bfr+\frac{d}{2}|+a}.
\end{align}
This idea can be extended by defining additional wells to move from an atomistic view to a material or crystalline structure in one dimension.

All data used to train the presented models were generated using external potentials defined in this manner. The electronic densities and KS eigenfunctions were generated using the iDEA code~\cite{hodgsonExactTimedependentDensityfunctional2013a, IDEAInteractingDynamic2023}. 
For the one-dimensional atom of varying charge, $Z$ was varied between the values of two and six in steps of $0.1$, resulting in a dataset with 41 individual data points which was split into a train, test, and validation set. These sets had 32, 5, and 4 individual data points each, respectively. 
For the one-dimensional diatomic molecule, $Z_1$, $Z_2$, and $d$ were all varied within the range of one to five, in steps of $0.5$. All combinations of the three variables were generated, resulting in a data set with 729 data points. The dataset was split such that the training set had 590 points, the test set had 73, and the validation set had the remaining 66.

\subsection{Physics-informed neural networks}
As a subset of neural networks, PINNs use equations and conditions of the underlying physics to enable a neural network to predict physically relevant information. Consider a general partial differential equation of the form
\begin{align}
    u(x,t) + \mathcal{N}[u;\lambda] = 0\,,
\end{align}
where $u$ is the solution, $x \in \Omega$ a $D$-dimensional spatial coordinate with $\Omega$ a subset of $\mathbb{R}^D$, $t \in [0,T]$ the time variable, $\mathcal{N}$ a nonlinear operator, and $\lambda$ the set of parameters that describe a physical system. If a neural network is used to represent the solution $u$, as $\tilde{u}$, this gives rise to
\begin{align}
    \tilde{u}(x,t) + \mathcal{N}[\tilde{u};\lambda] := f(x,t)\ .
\end{align}
The function $f$ must only equal zero when the network accurately represents the solution to the differential equation for the given parameters $\lambda$. If trained on a wide set of physical data, it should be able to predict solutions to the differential equation within that set more generally.

Based on \Eqref{KSeqn}, we derive the following model and loss functions within the framework of PINNs:
\begin{align}
    f(\bfr) = \epsilon_i ~ \phi_i(\bfr) + \frac{\nabla^2 ~ \phi_i(\bfr) }{2} - v_{s}(\bfr) ~ \phi_i(\bfr)
\end{align}
\begin{align}
L\pde = \mathrm{}{MSE}(f,0) = \frac{1}{N_f} \sum_{i=1}^{N_f}\left| f(x_f^i) \right|^2\,,
\end{align}
where the loss function $L\pde$ is computed in terms of a mean squared error (MSE) $\mathrm{}{MSE}(f,0)$,  with $x_f^i$ denoting one of the $N_f$ collocation points where the solution is provided.

Here, our focus is to invert the KS equation, wherein we have the set of solutions, $\{\phi_i\}$, and our goal is to get the network to predict the parameters, $v\s$, that lead to an appropriate set of solutions. If the network can correctly predict $v\s$, then the function $f$ should return a value of zero. What distinguishes this approach from conventional data-driven ML is that the network does not receive the solution set as input for prediction. Instead, it receives only the weighted sum of the solutions in the form of the electronic density calculated using \Eqref{denseqn}.

\begin{figure*}
    \centering
    \includegraphics[scale=.35]{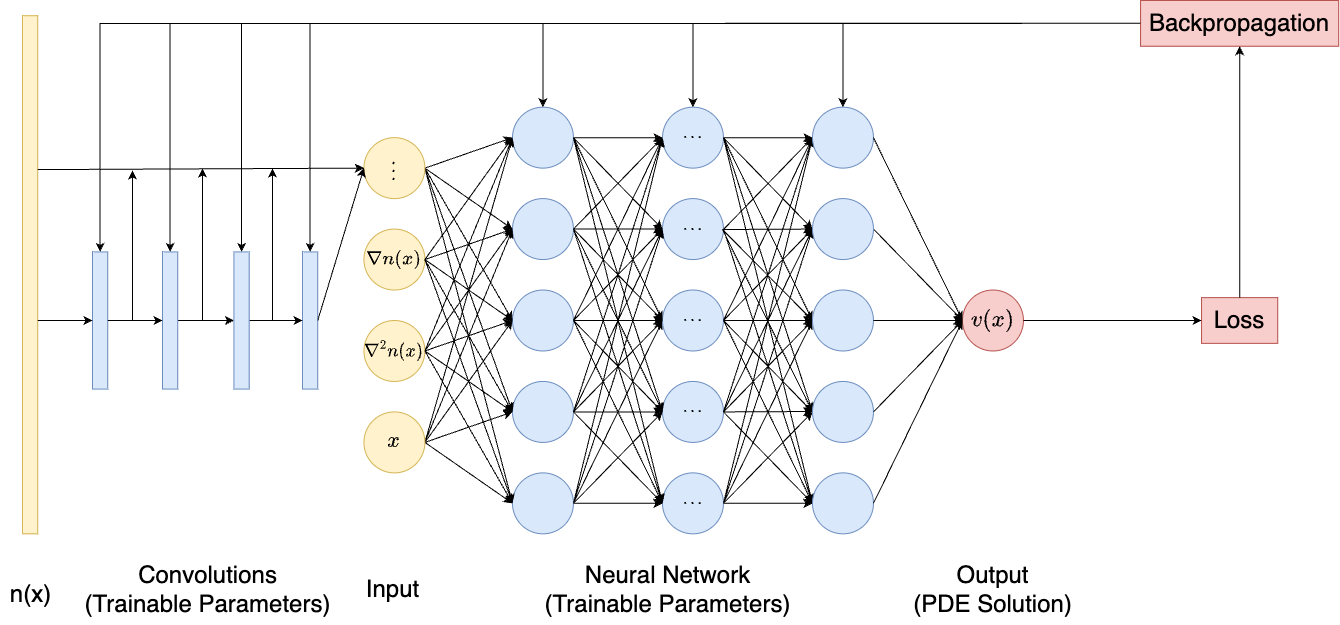}
    \caption{PINN architecture for predicting the KS potential based on an input electronic density.}
    \label{fig:PINN_arch}
\end{figure*}

In Figure~\ref{fig:PINN_arch}, we illustrate our implementation of a PINN that predicts the KS potential for an input density. The network takes the density as an input and processes it through a series of convolutions. The output of each convolution as well as the initial density is used as an input into a dense network. The network is then asked to predict the KS potential on each point in space given this set of features, as well as the first and second derivative of the density.

Convolutional neural networks are a common ML technique for image recognition \cite{KSH17,SK15,HZR16,SLJ15}. They observe both local and non-local features in input data to predict what is present in an image. 
The convolution of two functions $f$ and $g$ is defined as
\begin{align}
    (f*g)(t) = \int f(\tau)~g(t-\tau)~d\tau
\end{align}
and can be described as sum of the overlap of two functions at each point in $t$. 
In our convolutional PINN, this is accomplished by using a set of convolutions in series and in parallel. Back propagation is used to learn functions $g$ that emphasize input features crucial for predicting a class. Increasing the number of convolutions in series results in the passing of additional non-local information to the features.

It is widely known\cite{HK64,LYP88} that the XC potential is a nonlocal functional of the electronic density. Consequently, training a network to correctly predict an XC potential based only on the electronic density at a single point in space is ill-advised. Our PINN model uses convolutions to account for some of the non-locality of the XC potential. Each step through the convolutions indicates a larger regime of non-locality that has been included in its output features. 

\subsection{Fourier neural operators}  \label{FNO_intro}
Neural operators (NOs) \cite{luDeepONetLearningNonlinear2019, anandkumarNeuralOperatorGraph2020} are an extension of neural networks that take functions as inputs and return functions as outputs. While traditional neural networks map finite-dimensional vectors to finite-dimensional vectors, neural operators work on infinite-dimensional function spaces. A neural network can be defined as a mapping \( f: \mathbb{R}^n \rightarrow \mathbb{R}^m \), whereas a neural operator maps functions to functions, i.e., \( G: \mathcal{A} \rightarrow \mathcal{U} \), where \(\mathcal{A}\) and \(\mathcal{U}\) are function spaces. Given that there exists a non-linear map $\mathcal{G}^{\dagger}: \mathcal{A} \rightarrow \mathcal{U}$, our goal is to construct a neural operator that approximates this map, i.e., \( \mathcal{G_\theta} \approx \mathcal{G}^{\dagger} \), where $\mathcal{G}_\theta: \mathcal{A} \rightarrow \mathcal{U}$ with parameters $\theta$ in finite dimensional space $\mathbb{R}^p$.

The neural operator is trained on point-wise (in function space) observations of the form $ \{ (a_i, u_i) \}_{i=1}^N $, where $ \left( a_i \in \mathcal{A}; a_i(x) \in \mathbb{R}, x \in \mathbb{R} \right)$, $ \left( u_i \in \mathcal{U}; u_i(x) \in \mathbb{R}, x \in \mathbb{R} \right)$ and $u_i = \mathcal{G^{\dagger}} (a_i) $. The goal is to find the parameters $\theta^*$ that minimize the loss
\begin{align}
    \theta^* = \min _{\theta \in \mathbb{R}^p} \frac{1}{N} \sum_{i=1}^N\left\|u^{(i)}-\mathcal{G}_\theta(a_i)\right\|_{\mathcal{U}}^2\ .
\end{align}
Analogous to the neural networks being defined as multi-layer compositions of linear and non-linear operations, we can define a neural operator $\mathcal{G}_\theta(a)$ as multi-layer compositions of linear and non-linear operators acting directly in function space:
\begin{align}
\mathcal{G}_\theta(a) = Q(v_L(v_{L-1}(\dots v_1(P(a)))))\,,
\label{eq:no_definition}
\end{align}
where the layer $v_{l+1}$ is defined as
\begin{align}
    v_{l+1}(x)=\sigma_{(l+1)}\left(W_l v_l(x)+\left(\mathcal{K}_l(a ; \lambda) v_l\right)(x)\right)\ .
    \label{eq:no_layer}
\end{align}
Here, the non-local kernel integral operator $\mathcal{K}_l$ is defined as
\begin{align}
    \left(\mathcal{K}_l(a ; \lambda) v_l\right)(x)=\int_{\Omega_{l}} \kappa_{l}(x, y,a(x), a(y)) v_l(y) d y
    \label{eq:no_kernel}
\end{align}
over the domain $\Omega_l$. 

The kernel function $\kappa_l(x,y,a(x), a(y))$ is a function of $x,y$, and the input functions $a(x)$ and $a(y)$. The kernel function, parametrized by $\lambda$, and the local linear operator $W_l$, a matrix of weights, are learned during training. $\sigma_{(l+1)}$ is a local non-linear activation function. The local operator $P(a)$ is a pre-processing operator that transforms the input function $a$ into a higher dimensional representation $v_0$, which is the input to the first layer. The local operator $Q(v_L)$ is a post-processing operator that projects the output of the last layer $v_L$ back into $\mathcal{U}$. 

Several methods can be used to approximate the global integral calculation in each layer, such as graph neural networks with Monte-Carlo sampling \cite{anandkumarNeuralOperatorGraph2020} and multipole graph neural operators. ~\cite{liMultipoleGraphNeural2020} These approximations reduce the computational complexity and allow for scalable solutions \cite{kovachkiNeuralOperatorLearning2023a}.

Fourier neural operators (FNOs)\cite{liFourierNeuralOperator2020} leverage the Fourier transform to work in the frequency domain. Notice that by using a kernel $\kappa(x-y)$, \Eqref{eq:no_kernel} is a convolution operator, and by using the convolution theorem, \Eqref{eq:no_kernel} becomes
\begin{align}
    \left(\mathcal{K}_l(a ; \phi) v_l\right)(x)=\mathcal{F}^{-1}\left(\mathcal{F}\left(\kappa_l\right) \cdot \mathcal{F}\left(v_l\right)\right)(x) \\
    \mathcal{F}(x) = \sum_{n=0}^{N-1}~x_n \cdot e^{-\frac{i2\pi}{N}kn} = X_k \\
    \mathcal{F}^{-1}(X) = \frac{1}{N}\sum_{n=0}^{N-1}~X_k \cdot e^{\frac{i2\pi}{N}kn} = x_n,
\end{align}
where $\mathcal{F}$ is the Fourier transform and $\mathcal{F}^{-1}$ is its inverse. The kernel $\kappa_l$ is parametrized in Fourier space. Here the parameterization is finite-dimensional determined by the number of Fourier modes $k_{max}$.
By working in frequency domain, the kernels can be computed efficiently using the Fast-Fourier Tranform (FFT) and the model can capture global patterns more effectively.

\begin{figure*}
    \centering
    \includegraphics[scale=0.9]{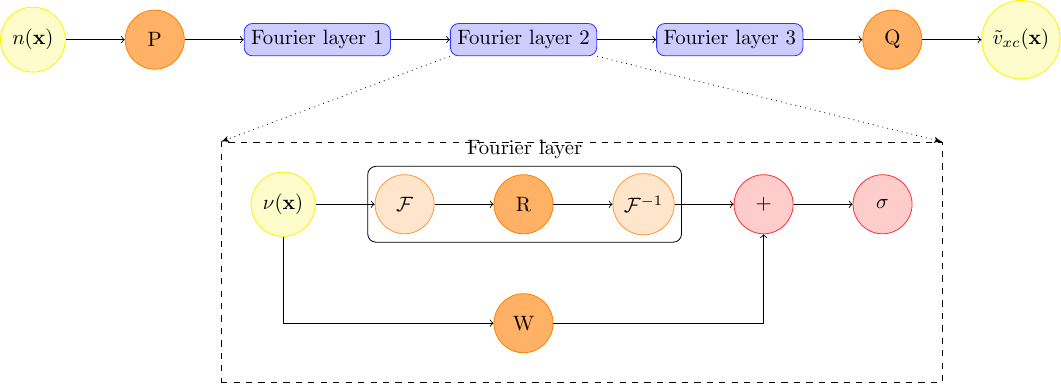}
    \caption{FNO architecture for predicting the KS potential based on an input electronic density.}
    \label{fig:FNO_arch}
\end{figure*}

A distinction has to be made between the solving approach (PINNs) and the learning approach (NOs) for PDE problems. PINNs are useful for obtaining numerical solutions for specific initial and boundary conditions and PDE parametrizations. In contrast, learning a PDE via neural operators aims to learn the underlying operator itself, enabling the prediction of solutions for various conditions without re-solving the PDE. PINNs can be used to solve PDEs if the PDEs are defined, even in the absence of training data, while NOs can be used to approximate undefined PDEs with training data. NOs are resolution-invariant and can also be used for zero-shot super-resolution \cite{liFourierNeuralOperator2020}.  A rigorous mathematical definition and analysis for NOs is provided in \cite{kovachkiNeuralOperatorLearning2023a}. The strengths of FNOs in this context lie in their inherent capability to capture non-local dependencies and their resolution invariance.

Given the non-local dependence of the XC potential on the electronic density, the non-locality of FNO layers make them well-suited for the inverse problem in KS-DFT. The input in this case is a grid of electronic density values and the output is the XC potential at the corresponding spatial grid points. The loss function is simply the MSE between true and predicted XC potential: 
\begin{align}
L_{FNO} = \mathrm{}{MSE}(v\xc, \tilde{v}\xc) = \frac{1}{N} \sum_{i=1}^{N}\left| v^{(i)}\xc - \tilde{v}^{(i)}\xc\right|^2\,
\end{align}
where $\tilde{v}\xc$ denotes the predicted XC potential for system $i$.
The architecture we implemented is shown in Figure~\ref{fig:FNO_arch}.

\subsection{Similarity Measures}
We evaluate the complexity of the datasets using two different similarity measures, namely the cosine similarity and Euclidean distance. The cosine similarity between two vectors $\mathbf{X}$ and $\mathbf{Y}$ in $\mathbb{R}^D$ is defined as
\begin{align}
    \textnormal{cos}(\theta) = \frac{\mathbf{X} \cdot \mathbf{Y}}{\Vert \mathbf{X} \Vert \Vert \mathbf{Y} \Vert},
\end{align}
which returns values in the range $[-1,1]$, namely $-1$ if the vectors point in opposite directions, $1$ if they point in the same direction, and 0 if they are orthogonal to each other. The Euclidean distance between the same vectors is defined as
\begin{align}
    D( \mathbf{X}, \mathbf{Y} ) = \Vert \mathbf{X} - \mathbf{Y} \Vert.
\end{align}
This measure has a lower bound of zero but no upper limit. The value returned is 0 only if the vectors $\mathbf{X}$ and $\mathbf{Y}$ are the same vector. Then, as two vectors continue to move further apart, the value moves further away from zero. The Euclidean distance includes some details related to the cosine similarity, but it also describes the difference in magnitude of the values found in the two vectors. This is a significant piece of information that the cosine similarity measure misses. Both of these measures provide insight into the complexity of each dataset and the network architecture's capacity to work on each of them.

\begin{figure}[htbp]
    \centering
    \includegraphics[scale=.5]{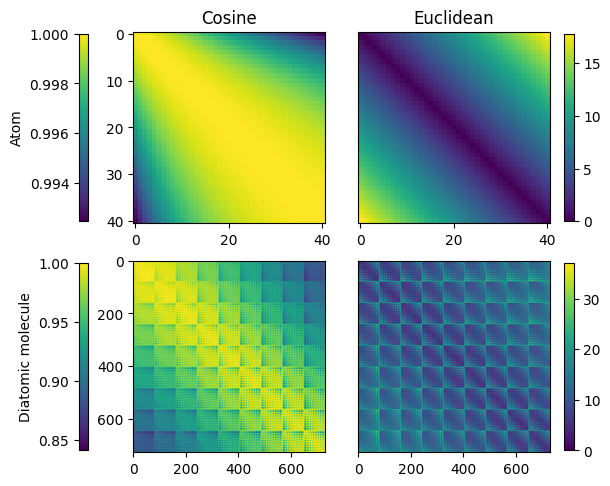}
    \caption{Similarity measures $-$ cosine similarity (left) and Euclidean distance (right) $-$ of the two model systems. The top panel shows the similarity measures for the one-dimensional atom with varying charge, while the bottom panel shows the similarity measures for the one-dimensional diatomic molecule with varying charge and bond length. The similarity measures illustrate the greater degree of complexity in the molecular model system. The axes are labeled by the system index.}
    \label{fig:sim_single}
\end{figure}

We illustrate the similarity for both the atomic and the molecular model systems in Figure~\ref{fig:sim_single}. 
As expected, the diatomic system is more complex than the atomic system, which is evident from the broader range of values in the diatomic case, versus the atomic case, for both cosine similarity and Euclidean distance measures. In the cosine similarity plots for the diatomic system, a block pattern emerges. The largest blocks shift with a change in the bond length. Descending to the bottom or moving to the right of the plot corresponds to an increase in the separation between two atomic centers. The greatest dissimilarity is noted at the two extremes of separation. Interestingly, at larger separations, the vectors show slightly less self-similarity compared to those at lower separations, as indicated by the darker shading of the bottom right block compared to the top left. The Euclidean distance plot yields similar trends, although it does not show this distinction within varying separation distances.

\section{Results}
We next present our findings for the two PIML approaches, comparing results using the PINN model as well as the FNO model. We trained both models on the same datasets and evaluate the model performance via the accuracy of the returned eigenvalues when solving the KS equations with the predicted XC potentials. We assess the ability of both PIML approaches as inverters for the KS equations in terms of the two aforemorentioned model systems: the one-dimensional atom with varying charge and the diatomic molecule with varying charges and bond length. Both the PINN model and the FNO model were trained on model systems with a grid resolution of 301 in real space. The performance was measured for systems with resolutions of 301 and 501, as well as for 301-grid systems with parameters outside the training range (extrapolation).

In all model systems, we consider the first six eigenstates, where each model is occupied with two same-spin fermions. Additionally, we consider four virtual KS orbitals.  The set of occupied and virtual eigenstates defines the shape of the KS potential. The lowest-energy eigenvalue indicates how well networks predict the cusp of the atoms, while the highest value signifies how well they predict the tail of the potential. 

We evaluate each model for ten neural networks with different random seeds. The error is quantified using the mean absolute error (MAE) and the mean absolute percentage error (MAPE) of the eigenvalues defined as
\begin{align}
\text{MAE}_{i,j} &= \frac{1}{N} \sum_{n=1}^{N} \left| \tilde{\epsilon}_{i,j,n} - \epsilon_{i,j} \right|\\    
\text{MAPE}_{i,j} &= \frac{1}{N} \sum_{n=1}^{N} \left| \frac{\tilde{\epsilon}_{i,j,n} - \epsilon_{i,j}}{\epsilon_{i,j}} \right| \times 100\%\,,
\end{align}
where $\tilde{\epsilon}_{i,j,n}$ denotes the predicted eigenvalues and each predicted eigenvalue is indexed by $i,j,n$, with $i$ denoting the level index, $j$ the system index, and $n$ the network index, with each neural network having a different random seed.

\subsection{Inversions with physics-informed neural networks}

\subsubsection{Atom with varying charge~~}
\begin{figure}[htbp]
    \centering
    \includegraphics[scale=.5]{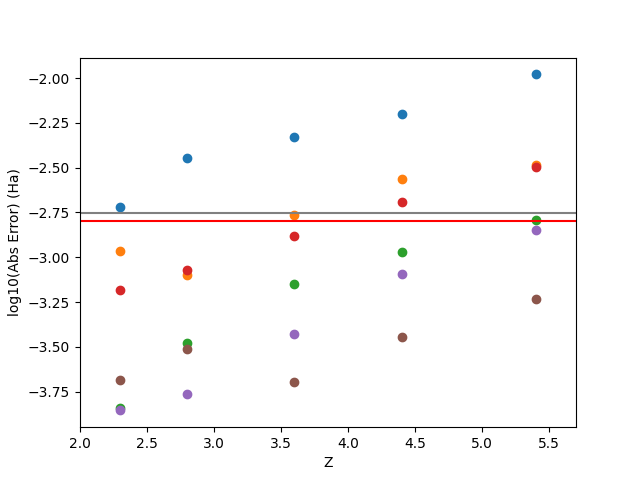}
    \caption{
    MAE of eigenvalues predicted by the PINN model on a logarithmic scale for the one-dimensional atom. 
    The MAE is averaged over 10 network initializations. The red horizontal line denotes chemical accuracy, while the gray horizontal line depicts the MAE averaged over all test systems. The colors label the eigenstates in increasing order as follows: blue, orange, red, green, purple, and brown.}
    \label{fig:pinn_atom_mae}
\end{figure}

Figure~\ref{fig:pinn_atom_mae} illustrates the performance of the PINN model for a one-dimensional atom with a variable parameter $Z$ corresponding to its charge, plotting the MAE for a set of fixed charges. Each data point represents the mean of ten neural network predictions that were generated using varying network initializations and random seeds. The data points are color-coded to indicate ground and excited states in the order of blue, orange, red, green, purple, and brown. This order is maintained in the following discussion of the results. 

The performance of the PINN model is best when the well charge $Z$ is small. The error systematically increases with the charge, particularly for the lowest-lying eigenvalue. The overall MAE averaged over the test systems is 1.77e-3 Ha (denoted by a gray horizontal line in Figure~\ref{fig:pinn_atom_mae}). The PINN model is competitive with the chemical accuracy criterion of being within 1 kcal/mol or 0.0016 Ha, as marked by the red horizontal line in our figures. 

\begin{figure}[htbp]
    \centering
    \includegraphics[scale=0.5]{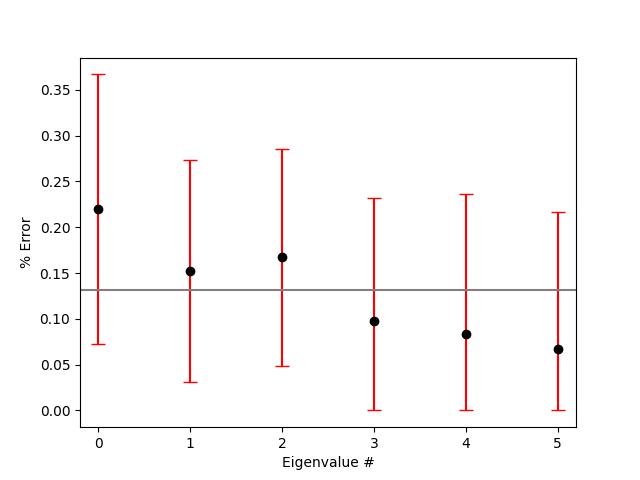}
    \caption{The MAPE resolved for the first six eigenvalues of the one-dimensional atom with varying charges predicted by the PINN model. The red error bars denote three standard deviations away from the mean, while the gray horizontal line depicts the MAPE averaged over all eigenvalues.}
    \label{fig:pinn_atom_mape}
\end{figure}

We have also displayed the the MAPE of the individual eigenvalues in \ref{fig:pinn_atom_mape}. The greatest error is in the first two eigenvalues of the systems, with both lying above the total average MAPE of around .13\%, while the next four lie below this threshold.

We have also explored the relationship between the performance of the PINN model and the resolution of the spatial grid. We suspect that the source of the errors observed in Figure~\ref{fig:pinn_atom_mae} is related to the resolution of the grid used to train the PINN model, which depends on the accuracy of the KS orbitals and their second derivatives. We anticipate that as the grid resolution improves, the accuracy of the second derivatives will also improve, particularly at the sharp cusp-like feature in the KS potential. Our findings indicate that errors diminish with the enhancement of grid resolution. This correlation presents a methodical approach for reducing the error in the PINN model. The improvements are reported in Table~\ref{tab:mean_errors}, which showcases the systematic enhancement in performance with increasing grid resolution (see Atom 501).

\subsubsection{Diatomic molecule with varying charges and bond length~~}
\begin{figure}[htbp]
    \centering
    \includegraphics[scale=.5]{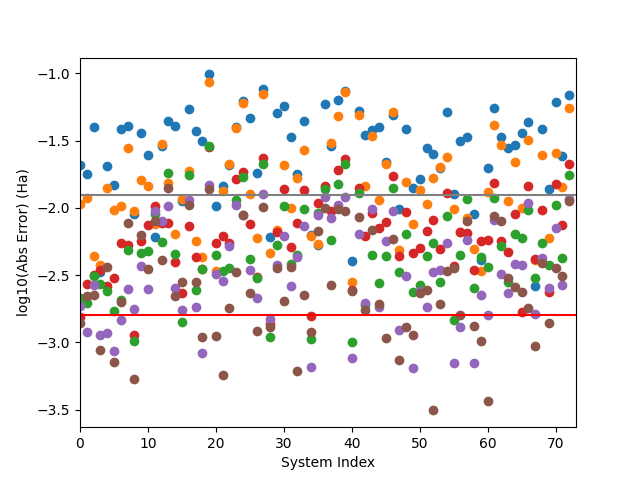}
    \caption{MAE of eigenvalues predicted by the PINN model on a logarithmic scale for the one-dimensional diatomic molecule. 
    The MAE is averaged over 10 network initializations. The red horizontal line denotes chemical accuracy, while the gray horizontal line depicts the MAE averaged over all test systems. The colors label the eigenstates in increasing order of energy as described in Figure~\ref{fig:pinn_atom_mae}.}
    \label{fig:pinn_molecule_mae}
\end{figure}

When switching to the diatomic molecule, the learning task becomes much more complex, as both the charges ($Z_1$ and $Z_2$) and the bond length ($d$) are variable parameters. This is exemplified in the roughly 10-fold increase in errors within the predicted eigenvalues. The MAE is shown in Figure~\ref{fig:pinn_molecule_mae}. Compared to the atomic test system, which had an overall MAE averaging around 1.77e-3 Ha, the overall MAE has increased to 1.25e-2 Ha (refer to the gray horizontal line in Figure~\ref{fig:pinn_molecule_mae}). In the atomic test system, the largest errors were observed in the ground, first, and second states of the KS spectrum. All eigenvalues exhibit some errors above the limit of chemical accuracy. This suggests that the PINN model has difficulty predicting the tails of the XC potential, but not to the same extent as placing and accurately estimating the well depth. The relative ordering of errors on the states remains consistent.

\begin{figure}[htbp]
    \centering
    \includegraphics[scale=0.5]{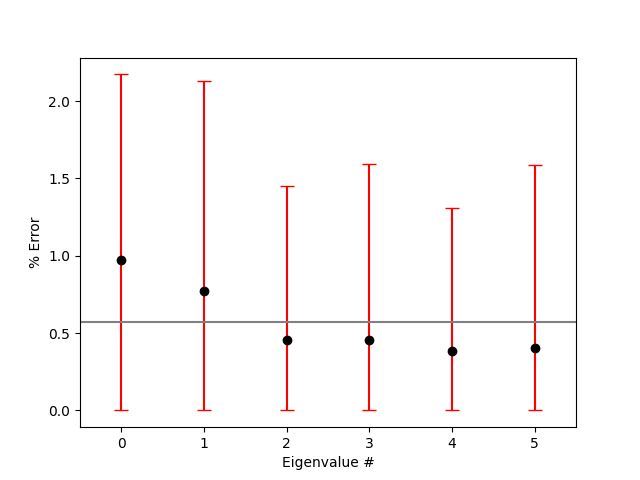}
    \caption{The MAPE resolved for the first six eigenvalues of the one-dimensional diatomic molecule with varying charges and bond length predicted by the PINN model. The red error bars denote three standard deviations away from the mean, while the gray horizontal line depicts the MAPE averaged over all eigenvalues.}
    \label{fig:pinn_molecule_mape}
\end{figure}

We have further broken down the error analysis into individual eigenvalues in Figure~\ref{fig:pinn_molecule_mape}. We display the MAPE for the first six eigenvalues, which exhibits the larger errors for the ground and first excited states observed in the MAEs plot. The overall MAPE averages around .575\% (denoted by a gray horizontal line in Figure~\ref{fig:pinn_molecule_mape}).

\subsubsection{Extrapolation for diatomic molecule~~}
Purely data-driven neural network approaches rapidly degrade in predictive performance when provided with data outside the training set. However, using physics-based approaches such as PINNs and also FNOs, higher extrapolation performance can be expected. Compared to data-driven learning, the neural network in the PINN model represents the solution to a given PDE, allowing accurate predictions within the specific scope of that PDE. We evaluate the PINN model's extrapolation ability by asking it to predict on a set of densities outside the training set.

\begin{figure}[htbp]
    \centering
    \includegraphics[scale=0.5]{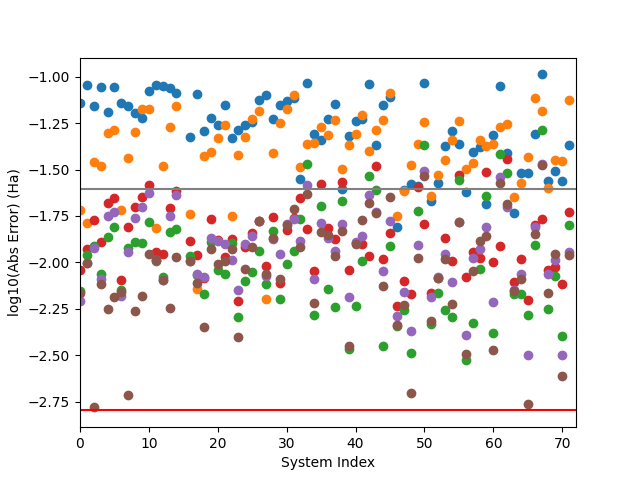}
    \caption{Extrapolation with the PINN model for the one-dimensional diatomic molecule with varying charges and bond length. The test set includes molecules with charges and bond length that are outside the parameter ranges of the training dataset.
    The MAE of the predicted eigenvalues is shown on a logarithmic scale. The MAE is averaged over 10 network initializations. The red horizontal line denotes chemical accuracy, while the gray horizontal line depicts the MAE averaged over all test systems. The colors labels the eigenstates in increasing order of energy as described in Figure~\ref{fig:pinn_atom_mae}.}
    \label{fig:pinn_molecule_mae_extra}
\end{figure}

\begin{figure}[htbp]
    \centering
    \includegraphics[scale=0.5]{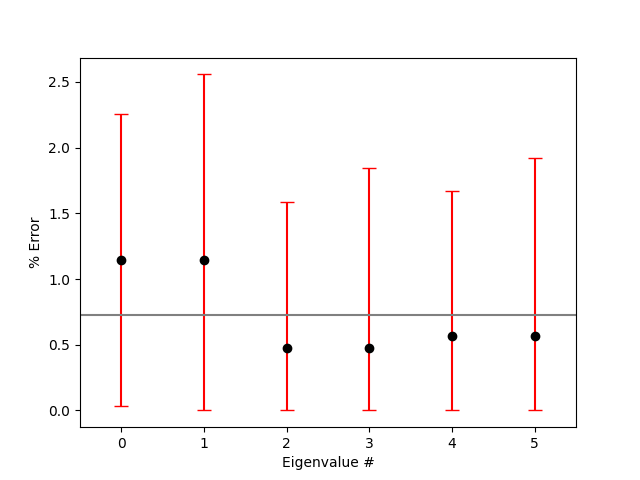}
    \caption{Extrapolation with the PINN model for the one-dimensional diatomic molecule. The MAPE is resolved for the first six eigenvalues. The red error bars denote three standard deviations away from the mean, while the gray horizontal line depicts the MAPE averaged over all eigenvalues.}
    \label{fig:pinn_molecule_mape_extra}
\end{figure}

The outcome of this analysis is illustrated in Figure~\ref{fig:pinn_molecule_mae_extra}, which shows the MAE. 
The errors have increased as anticipated, resulting in an overall MAE of 2.48e-2 Ha indicated by the gray horizontal line. The error has doubled in comparison to the interpolation task featured in Figure~\ref{fig:pinn_atom_mae}. While this is not a catastrophic failure, the higher error suggests the intricacy of extrapolation. The challenge of extrapolation is evident in Figure~\ref{fig:pinn_molecule_mape_extra}, where we illustrate the MAPE resolved over the first six eigenvalues with an overall MAPE of .73\%, denoted by a gray line. Although no significant or sudden rise in error is observed with increasing eigenvalues, the overall MAPE is noticeably higher, as is the standard deviation shown by the red error bars.

\subsection{Data-driven inversions with Fourier neural operators}
In contrast to the PINN model, the density-to-potential mapping data is explicitly used in the loss function to train the data-driven FNOs, as outlined in Section~\ref{FNO_intro}. The network takes the density grid as input and produces the corresponding potential grid as output. FNOs have several benefits, including learning the mapping in the functional space, which enables them to generalize effectively over varying input densities. Additionally, the network is resolution-invariant, allowing it to assess higher resolution grids than those present in the training dataset.

\subsubsection{Atom with varying charge~~}
\begin{figure}[htbp]
  \centering
  \includegraphics[scale=0.5]{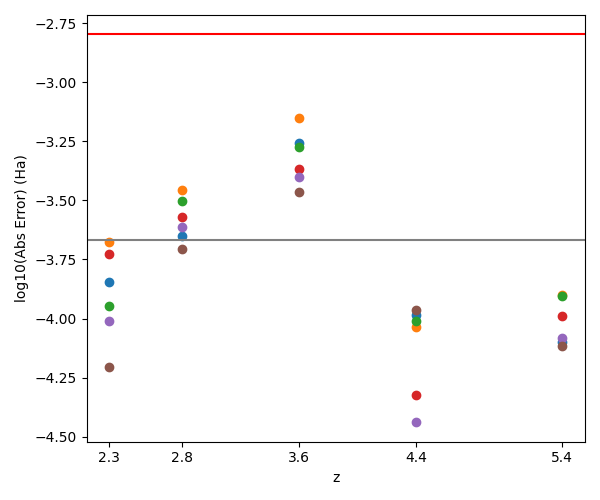}
  \caption{MAE of eigenvalues predicted by the FNO model on a logarithmic scale for the one-dimensional atom. The MAE is averaged over 10 network initializations. The red horizontal line denotes chemical accuracy, while the gray horizontal line depicts the MAE averaged over all test systems. The colors label the eigenstates in increasing order of energy as described in Figure~\ref{fig:pinn_atom_mae}.}
  \label{fig:fno_atom_mae}
\end{figure}

\begin{figure}[htbp]
  \centering
  \includegraphics[scale=0.5]{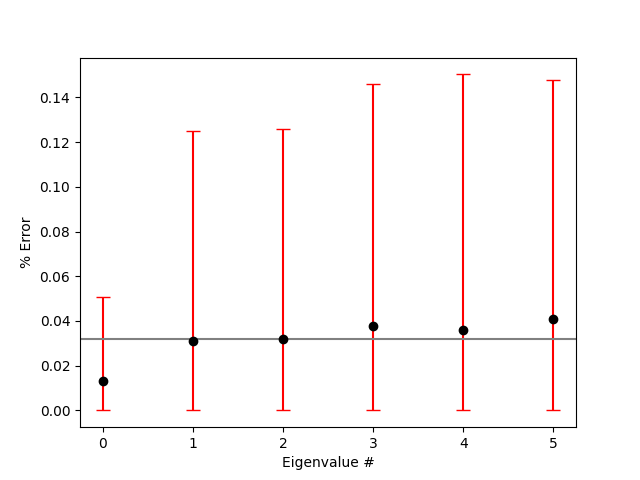}
  \caption{The MAPE resolved for the first six eigenvalues of the one-dimensional atom with varying charge predicted by the FNO model. The red error bars denote three standard deviations away from the mean, while the gray horizontal line depicts the MAPE averaged over all eigenvalues.}
  \label{fig:fno_atom_mape}
\end{figure}

In the case of the one-dimensional atom with varying charge $Z$, the FNO model demonstrates exceptional performance over the range of charge values. The error plots for the eigenvalues, as shown in Figure~\ref{fig:fno_atom_mae}, indicate that the predicted eigenvalues closely align with the true values. We resolve the error analysis over the spectrum of eigenvalues by showing the MAPE for the first six eigenvalues in Figure~\ref{fig:fno_atom_mape}. The red error bars denote three standard deviations away from the mean, demonstrating that the MAPE is within 0.15\% in this range of eigenvalues. The range of MAPEs is lowest for the ground state and shows a slight increasing trend with higher energy levels. This increase can be attributed to the magnitude of the eigenvalues decreasing as the energy level increase. Higher energy states, with their inherently lower magnitudes, offer a smaller margin for absolute error, thus demanding greater accuracy.  The model achieves this to a lesser degree, although the level of error is still well within the range of chemical accuracy. The overall MAE averaged over all model systems is 2.22e-04 Ha and the corresponding averaged MAPE over all eigenvalues is 3.18e-02 \% denoted by gray horizontal lines in Figures~\ref{fig:fno_atom_mae} and \ref{fig:fno_atom_mape}. This underscores the FNO model's ability to accurately capture the potential-density mapping.

\subsubsection{Diatomic molecule with varying charges and bond length~~}
\begin{figure}[htbp]
  \centering
  \includegraphics[scale=0.5]{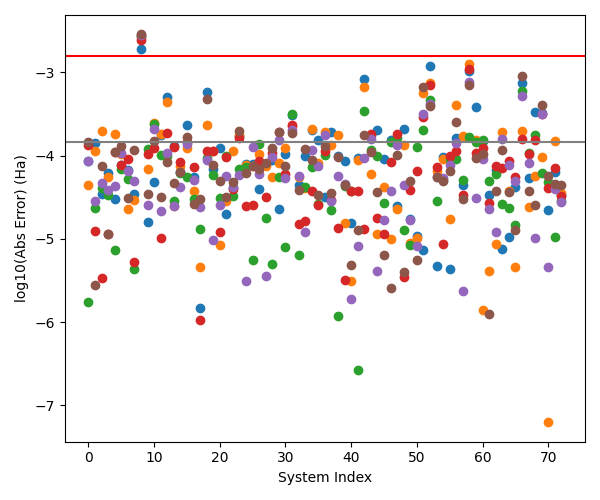}
  \caption{MAE of eigenvalues predicted by the FNO model on a logarithmic scale for the one-dimensional diatomic molecule. The MAE is averaged over 10 network initializations. The red horizontal line denotes chemical accuracy, while the gray horizontal line depicts the MAE averaged over all test systems. The colors label the eigenstates in increasing order of energy, as described in Figure~\ref{fig:pinn_atom_mae}.}
  \label{fig:fno_molecule_mae}
\end{figure}

\begin{figure}[htbp]
  \centering
  \includegraphics[scale=0.5]{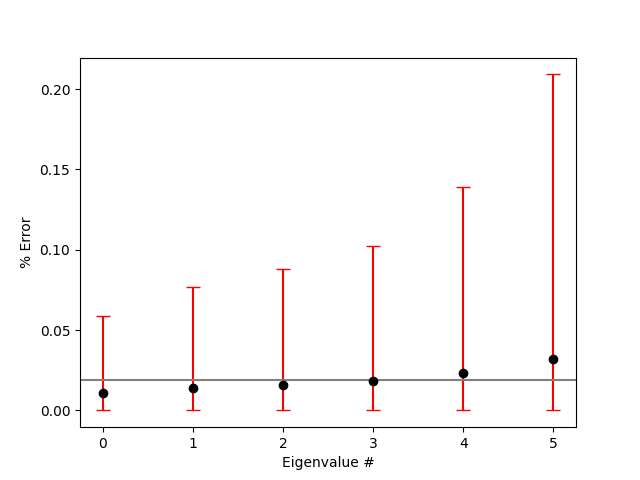}
    \caption{The MAPE resolved for the first six eigenvalues of the one-dimensional diatomic molecule with varying charges and bond length predicted by the FNO model. The red error bars denote three standard deviations away from the mean, while the gray horizontal line depicts the MAPE averaged over all eigenvalues.}
  \label{fig:fno_molecule_mape}
\end{figure}

After demonstrating improved performance over the PINNs model for the one-dimensional atom, we investigate the FNO model's abilities for the one-dimensional diatomic molecule. Variations in both charges ($Z_1$ and $Z_2$) and bond length $d$ pose a different set of challenges for the FNO model. As depicted in Figure~\ref{fig:fno_molecule_mae}, the plots for eigenvalue errors indicate a slightly higher margin of error in comparison to the atomic model system. The eigenvalue errors consistently remain within chemical accuracy, except for a particular outlier. Figure \ref{fig:fno_molecule_mape} shows the MAPE range for each eigenvalue. There is a clear trend of increasing MAPE range from the ground state to excited states. Nevertheless, the overall MAE of 2.27e-04 Ha and MAPE of 1.90e-02\% (denoted by gray horizontal lines) align with the results of the atomic model system, confirming the FNO model's adaptability and accuracy across this more challenging set of system parameters.

\subsubsection{Extrapolation for diatomic molecule~~}
\begin{figure}[htbp]
  \centering
  \includegraphics[scale=0.5]{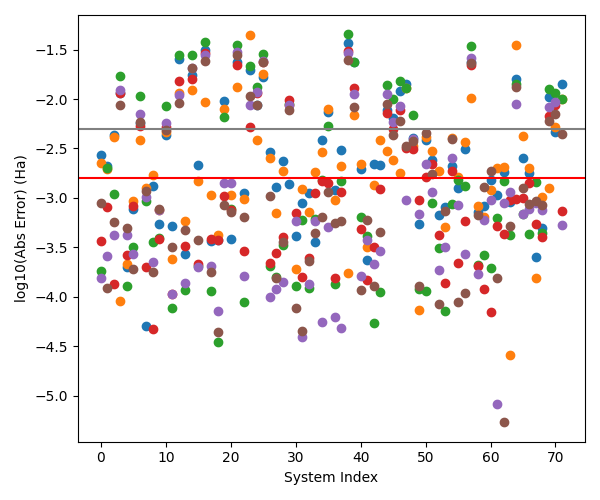}
  \caption{Extrapolation with the FNO model for the one-dimensional diatomic molecule. The test set includes molecules with charges and bond length that are outside the parameter ranges of the training dataset. 
  The MAE of the predicted eigenvalues is shown on a logarithmic scale. The MAE is averaged over 10 network initializations. The red horizontal line denotes chemical accuracy, while the gray horizontal line depicts the MAE averaged over all test systems. The colors labels the eigenstates in increasing order of energy as described in Figure~\ref{fig:pinn_atom_mae}.}
  \label{fig:fno_molecule_mae_extra}
\end{figure}

\begin{figure}[htbp]
  \centering
  \includegraphics[scale=0.5]{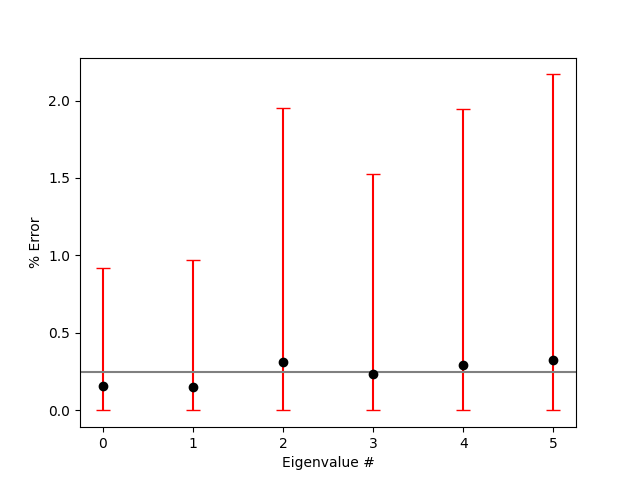}
  \caption{Extrapolation with the FNO model for the one-dimensional diatomic molecule. The MAPE is resolved for the first six eigenvalues. The red error bars denote three standard deviations away from the mean, while the gray horizontal line depicts the MAPE averaged over all eigenvalues.}
  \label{fig:fno_molecule_mape_extra}
\end{figure}

The true test of a predictive model lies in its ability to generalize beyond the parameters of the systems in the training dataset.  Figure~\ref{fig:fno_molecule_mae_extra} showcases the eigenvalue error plots for this dataset, which include systems with charges ($Z_1$ and $Z_2$) and bond lengths ($d$) that are outside the previous datasets. The errors recorded were marginally higher than those for the training dataset, and about half the eigenvalues are predicted within chemical accuracy. The percentage errors are within 2\%, as depicted in Figure \ref{fig:fno_molecule_mape_extra}. The FNO model yields overall an MAE of 2.27e-04 Ha and MAPE of 1.90e-02\% (denoted by gray horizontal lines). This demonstrates the FNO model's robustness and potential for broader applicability in solving the KS inversion problem.

\section{Discussion}
\begin{table*}
\centering
\begin{tabular}{ l c c c c }
\hline
\hline
\textbf{Model} & \textbf{Atom (301)} & \textbf{Atom (501)} & \textbf{Molecule (301)} & \textbf{Molecule (301 E)} \\ \hline
\multirow{4}{*}{PINN} & 0.0018 Ha & 0.0007 Ha & 0.0125 Ha & 0.0248 Ha \\
                      & \textit{(0.0110 Ha)} & \textit{(0.0034 Ha)} & \textit{(0.1000 Ha)} & \textit{(0.1200 Ha)}\\
                      & 0.13\% & 0.06\% & 0.58\% & 0.73\% \\
                      & \textit{(0.30 \%)} & \textit{(0.15\%)} & \textit{(2.37\%)} & \textit{(3.25\%)} \\ \hline
\multirow{4}{*}{FNO}  & 0.0002 Ha & 0.0002 Ha & 0.0002 Ha & 0.005 Ha \\
                      & \textit{(0.0009 Ha)} & \textit{(0.0010 Ha)} & \textit{(0.0051 Ha)} & \textit{(0.0584 Ha)}\\
                      & 0.03\% & 0.03\% & 0.02\% & 0.25\% \\
                      & \textit{(0.15\%)} & \textit{(0.19\%)} & \textit{(0.46\%)} & \textit{(3.51\%)}\\ \hline
\hline
\end{tabular}
\caption{Performance comparison of PINN and FNO models on both model systems, showing the MAE (in Ha) and MAPE (\%) and denoting the maximum absolute errors and the maximum percentage errors in braces.  The labels 301 and 501 indicate the grid resolutions and E denotes extrapolation.}
\label{tab:mean_errors}
\end{table*}
\begin{table*}
\centering
\begin{tabular}{ l c c c }
\hline
\hline
\textbf{Model} & \textbf{Atom (301)} & \textbf{Molecule (301)} & \textbf{Molecule (501)} \\
\hline
iDEA   & \(135 \pm 29\) & \(305 \pm 262\) & \(806 \pm 447\) \\
\hline
PINN  & \(0.0016 \pm 0.0011\) & \(0.0026 \pm 0.0013\) & N/A \\
\hline
FNO    & \(0.0022 \pm 0.0009\) & \(0.0020 \pm 0.0003\) & \(0.0037 \pm 0.0004\) \\
\hline
\hline
\end{tabular}
\caption{Computational timings for inversions (in seconds), comparing a conventional method (iDEA) with the PINN and FNO models.}
\label{tab:timings_inference}
\end{table*}
We summarize and contrast our results for the PINN and FNO models in Table~\ref{tab:mean_errors} for both atomic and molecular model systems. Our performance metrics are the MAE expressed in Hartree (Ha) and the MAPE. Overall, the FNO outperforms the PINN model, providing chemical accuracy (MAE $\le 0.0016$ Ha) for almost all model systems. 

For atoms, the PINN model shows an MAE of 0.0018 Ha, corresponding to a relative error of 0.13\%, while the FNO model shows an MAE of 0.0002 Ha, corresponding to a relative error of 0.03\%. The PINN model significantly improves its accuracy when the grid resolution is increased to 501, while the FNO model shows consistent accuracy over different grid resolutions.

When applied to molecules, the error of the PINN model increases significantly, especially for the extrapolation task, with an MAE reaching 0.0248 Ha and a relative error up to 0.73\%. The FNO model maintains a lower error rate and remains within the chemical accuracy except for the extrapolation task, where its MAE increases to 0.005 Ha and a relative error of 0.25\%. Despite the increase in error, the FNO model is still highly accurate. The values in parentheses represent additional metrics, namely the maximum absolute error and the maximum percentage error, which indicate the maximum spread of errors to evaluate the robustness of the models.

Having established the accuracy of the PIML methods, we turn to analyze the computational efficiency. To this end, Table~\ref{tab:timings_inference} shows the computational timings of the PINN and FNO models against a conventional inversion method (iDEA code) across the different model systems. Timing results are given in seconds, with standard deviations to indicate variability in computational performance. For atoms, the conventional inversion method shows a mean computation time of 135±29 seconds, which is significantly higher than both PINN and FNO models, which show mean computation times of 
0.0016±0.0011 seconds and 0.0022±0.0009 seconds, respectively. The difference is even more pronounced for the molecular systems, where the conventional method requires 305±262 seconds, while PINN and FNO remain in the millisecond range. In particular, for the molecular system with larger grid resolution (501 grid points), the computational time of the conventional method continues to increase, while FNO maintains an efficient computational evaluation in the millisecond range. These results highlight the substantial improvements in computational speed offered by the PINN and FNO models over conventional methods, with FNO showing slightly faster times and lower variability than PINN in the scenarios tested, suggesting a robustness that could be particularly beneficial for larger and more complex systems.

\section{Conclusion}
In summary, this work has demonstrated the potential of physics-informed machine learning (PIML) techniques, specifically physics-informed neural networks (PINNs) and Fourier neural operators (FNOs), to address the inverse problem in Kohn-Sham density functional theory (KS-DFT). The PINN model uses the underlying KS differential equation to predict the KS potential from given electron densities. FNOs, on the other hand, use data from known density-to-potential mappings. They exhibit superior performance in both accuracy and computational efficiency as PINNs across a range of grid resolutions and complexity of model systems. In particular, FNOs maintain chemical accuracy in most cases, underscoring their ability to handle the density-to-potential mapping effectively. The computational times for both PIML approaches significantly outperform conventional methods, offering a promising alternative for large-scale electronic structure inversions. 

A promising avenue for future work is to combine the strengths of both PINNs and FNOs. This could be realized by a hybrid model with a two-component loss term: one part based on the density-to-potential mapping characteristic of the FNO model, and another part derived from the underlying KS differential equation represented by the PINN model. During training, both the FNO and PINN components would be optimized simultaneously. This approach is expected to yield an efficient PIML model that not only exploits the data-driven robustness of the FNO component, but also requires less data and exhibits broader generalizability due to the PINN-derived loss. 

Our results indicate that the integration of machine learning, in particular PIML techniques, can significantly improve the efficiency and accuracy of electronic structure inversions. These advances offer promising prospects for elucidating the XC potential of materials from experimental electron density data and for advancing the development of XC functionals within density functional theory, providing a reliable computational tool for accurate density-to-potential inversions.

\section{Acknowledgments}
This work was partially supported by the Center for Advanced Systems Understanding (CASUS) which is financed by Germany’s Federal Ministry of Education and Research (BMBF) and by the Saxon state government out of the State budget approved by the Saxon State Parliament. 
Computations were performed on a Bull Cluster at the Center for Information Services and High-Performance Computing (ZIH) at Technische Universit\"at Dresden and on the cluster Hemera of the Helmholtz-Zentrum Dresden-Rossendorf (HZDR).
AC and KS acknowledge funding from the Helmholtz Association’s Initiative and Networking Fund through Helmholtz AI.

\section*{References}
\bibliographystyle{iopart-num}
\bibliography{bible.bib}

\end{document}